\documentclass[12pt]{article}
\usepackage[utf8]{inputenc}

\usepackage[margin=1in]{geometry}
\usepackage{booktabs} 
\usepackage[round,numbers,sort&compress]{natbib}

\usepackage{float}
\usepackage{graphicx}
\usepackage[colorlinks,citecolor=blue,urlcolor=blue,linkcolor=blue]{hyperref}
\usepackage{titling} 
\usepackage[superscript]{cite}
\usepackage{setspace}
\usepackage[document]{ragged2e}
\newcommand{\comment}[1]{}
\author{Peter B. Gilbert, Youyi Fong, Marco Carone}

\begin{document}
\section*{}
\normalsize
\textbf{Manuscript type:} Article
\vspace{0.4 in}
\begin{center} \Large{Assessment of Immune Correlates of Protection via Controlled Vaccine Efficacy and Controlled Risk}
\large
\vspace{0.3 in}

Running head: Controlled Vaccine Efficacy Immune CoP
\vspace{0.3 in}

Peter B. Gilbert\textsuperscript{a,b*}, Youyi Fong\textsuperscript{a,b}, Marco Carone\textsuperscript{b,a}
\end{center} 
\normalsize
\vspace{0.3 in}
\textsuperscript{a} Vaccine and Infectious Disease Division, Fred Hutchinson Cancer Research Center, Seattle, WA

\textsuperscript{b} Department of Biostatistics, University of Washington, Seattle, WA

\vspace{0.5 in}
*Correspondence: Peter B. Gilbert, Fred Hutchinson Cancer Research Center, 1100 Fairview Ave N, Seattle, WA, 98109. E-mail: pgilbert@scharp.org; Tel: (206) 667-7299.

\normalsize

\textbf{Conflicts of interest:} None declared.

\textbf{Sources of funding:} This work was supported by grants UM1AI068635 and R37AI054165 from the National Institute of Allergy and Infectious Diseases of the National Institutes of Health to Peter B. Gilbert.

\textbf{Data availability:} The CYD14 and CYD15 data are available upon request to the sponsor of the studies, Sanofi Pasteur.  The computing code used to implement the methods as illustrated in the CYD14 and CYD15 applications is available at the Github repository youyifong/CoPcontrolledVE. 

\textbf{Acknowledgments:} We thank the participants, investigators, and sponsor of the CYD14 and CYD15 trials, and Lindsay Carpp for editing contributions.

\newpage

\section*{Abstract}
\normalsize

Immune correlates of protection (CoPs) are immunologic biomarkers accepted as a surrogate for an infectious disease clinical endpoint and thus can be used for traditional or provisional vaccine approval.  
To study CoPs in randomized, placebo-controlled trials, correlates of risk (CoRs) are first assessed in vaccine recipients.  This analysis does not assess causation, as a CoR may fail to be a CoP.
We propose a causal CoP analysis that estimates the controlled vaccine efficacy
curve across biomarker levels $s$, $CVE(s)$, equal to one minus the ratio of the controlled-risk curve $r_C(s)$ at $s$ and placebo risk, where $r_C(s)$ is causal risk if all participants are assigned vaccine and the biomarker is set to $s$.
The criterion for a useful CoP is wide variability of $CVE(s)$ in $s$.
Moreover, estimation of $r_C(s)$ is of interest in itself, especially in studies without a placebo arm. 
For estimation of $r_C(s)$, measured confounders can be adjusted for by 
any regression method that accommodates missing biomarkers, to which we add 
sensitivity analysis to quantify robustness of CoP evidence to unmeasured confounding.  
Application to two harmonized phase 3 trials supports that 50\% neutralizing antibody titer has value as a controlled vaccine efficacy CoP for virologically confirmed dengue (VCD): in CYD14 the point estimate (95\% confidence interval) for $CVE(s)$ accounting for measured confounders and building in conservative margin for unmeasured confounding increases from 
29.6\% (95\% CI  3.5--45.9)
 at titer 1:36 to 78.5\% (95\% CI 67.9--86.8)
 at titer 1:1200; these estimates are
 17.4\% (95\% CI -14.4-- 36.5)
 and 84.5\% (95\% CI 79.6--89.1)
 for CYD15.
\comment{
\textbf{Conclusions:}
We recommend augmenting CoR analysis of vaccine recipients with controlled-risk CoP analysis that quantifies robustness of CoP evidence via sensitivity analysis. Additionally, in placebo-controlled trials controlled vaccine efficacy curve analysis provides interpretable CoP evaluation.
}

Key words: controlled effects causal inference, COVID-19, dengue vaccine efficacy, E-value, immune correlate of protection, sensitivity analysis.

\section{Introduction}
Safe and effective vaccines to prevent SARS-CoV-2 acquisition and COVID-19 disease are needed to curtail the COVID-19 pandemic.  Approval of a vaccine requires demonstration that the vaccine confers a favorable benefit-to-risk profile in reducing clinically significant endpoints, usually established through phase 3 randomized, placebo-controlled, vaccine efficacy trials. 
\cite{polack2020safetyshort,Badenetal2020short,sadoff2021safety} 
Where SARS-CoV-2 vaccines are approved and widely locally available, placebo arms in future efficacy trials will likely be infeasible.\cite{Follmannetal2020} Thus, there is a need for alternative approaches to approving SARS-CoV-2 vaccines, such as surrogate endpoint trials that use as primary endpoint an antibody response biomarker measured post-vaccination.  Effectiveness of this approach would require that the biomarker is measured using a validated assay and has been ``scientifically well established to reliably predict clinical benefit" (traditional approval) or be 
``reasonably likely to predict clinical benefit"\cite{FlemingPowers2012,FDASIA} (accelerated approval). 

Immunologic surrogate endpoints based on binding or functional antibody assays have been accepted by regulatory agencies for many licensed vaccines. \cite{Plotkin2008,Plotkin2010,PlotkinGilbert2018}  Acceptance has been based on evidence from a variety of sources, including statistical analysis of phase 3 vaccine efficacy trials, natural history studies, 
vaccine challenge studies in animals\cite{mason1973yellow,van2011correlation,beasley2016first}
and in humans,\cite{hobson1972role,chen2016single} and passive monoclonal antibody transfer studies. 
 Phase 3 trials constitute one of the most important sources of evidence, because they rigorously characterize the level of vaccine efficacy, and samples from breakthrough infection or disease cases can be analyzed and contrasted with samples from non-cases to infer immune correlates of risk (CoRs) and immune correlates of protection (CoPs).\cite{Qinetal2007} By CoP, we mean an immunologic biomarker statistically related to protection in some fashion that has been accepted for use in either  accelerated or traditional approval.\cite{PlotkinGilbert2018}
  Establishing a CoP typically requires multiple phase 3 placebo-controlled trials,
   with the placebo arm needed to enable assessment of criteria for a valid CoP within various statistical frameworks, including surrogate endpoint evaluation,\cite{Prentice1989,Freedman1992,Molenberghsetal2008} principal stratification vaccine efficacy moderation evaluation,\cite{GilbertGabrieletal2014,Moodieetal2018} mediation evaluation \cite{Cowlingetal2019}, stochastic interventional vaccine efficacy evaluation,\cite{Hejazietal2020} and meta-analysis.\cite{GabrielDanielsHalloran2016} 

This manuscript has two objectives. First, to propose a new causal inference approach to CoP evaluation based on randomized, placebo-controlled trials: controlled vaccine efficacy analysis. 
This approach is essentially based on estimation of the controlled causal effect of the immunologic biomarker in vaccine recipients on outcome risk, and is thus closely
linked to the second objective, to propose a causal inference approach to CoP evaluation based on analysis of the vaccine arm only.  The second objective is especially relevant for studies without a placebo arm. 
  As a case in point, several ongoing SARS-CoV-2 placebo-controlled vaccine efficacy trials have crossed over placebo recipients to the vaccine arm,\cite{polack2020safetyshort,Badenetal2020short,sadoff2021safety} precluding the study of CoPs against longer-term endpoints via methods requiring a placebo arm.
 These studies follow large numbers of vaccine recipients for SARS-CoV-2 acquisition and disease outcomes, providing the requisite data for vaccine-arm only CoP analysis. 
 
\section{Definition of a controlled vaccine efficacy CoP and a controlled risk CoP}

Based on analysis of a vaccinated group in a phase 3 trial or in a post-approval trial, regression methods (e.g. logistic or Cox regression accounting for case-cohort or case-control biomarker sampling\cite{Chan2002,LiParnesChan2013,Moodieetal2018}) may be used to identify CoRs, i.e., immunologic biomarkers measured from vaccine recipients associated with subsequent occurrence of a clinical endpoint of interest.\cite{mehrotra2020clinical,Qinetal2007} 
However, a CoR may fail to be a CoP, because the association parameter (e.g., hazard ratio) may not reflect a causal relationship.\cite{vanderWeele2013}  
Consequently, while identification of a CoR is an important step toward validating a CoP, it is itself insufficient.\cite{Prentice1989,FlemingDeMets1996}  We address this challenge in two steps. First, we define two causal effect parameters whose interpretations provide criteria for a CoP. Second, acknowledging that estimation of these two parameters requires the absence of unmeasured confounders, we develop a conservative estimation strategy that formally accounts for potential violations of this assumption. 

Let $A=1$ indicate assignment to vaccine, and if the study included a placebo, let $A=0$ indicate assignment to placebo.  
Let $S$ be an immunologic biomarker measured at a given post-vaccination visit, e.g. the Day 57 visit in the Moderna COVE phase 3 trial.\cite{Badenetal2020short} Let $Y$ be the indicator of occurrence of the clinical endpoint of interest after Day 57 during some fixed period of follow-up.  Let $M$ be the indicator that $S$ is measured; typically, $M=1$ for all outcome cases (with $Y=1$) with available samples at Day 57 as well as for a random sample of all enrolled participants (case-cohort design \cite{Prentice1986}) or of all non-cases (case-control design \cite{BreslowHolubkov1997}). Let $X$ be a vector of baseline covariates, including potential predictors of $Y$.  Let $O = (X,Y,M,MS)$ be the observed data unit for a participant, where $M S$ emphasizes that $S$ is only measured if $M=1$.

We define the two causal CoPs in terms of the causal parameter $r_C(s)=P\{Y(1,s)=1\}$, the probability of outcome occurrence for the counterfactual scenario in which all trial participants are assigned to vaccination and the immunologic biomarker is set to level $S=s$. We refer to the graph of $r_C(s)$ versus $s$ as the `controlled risk' curve since $r_C(s)$ forms the basis of a controlled effects parameter in the causal mediation literature.\cite{RobinsGreenland1992,Pearl2001}  We define $S$ to be a {\it controlled risk CoP} if $r_C(s)$ is monotone non-increasing in $s$ with $r_C(s) > r_C(s')$ for at least some $s < s'$. Point and confidence interval estimates of the graph of $r_C(s)$ versus $s$ describe the strength and nature of the CoP.

Building on a controlled risk CoP, for a placebo-controlled trial, we define the controlled vaccine efficacy curve \[\mathit{CVE}(s) = 1 - \frac{r_C(s)}{P\{Y(0)=1\}} = 1 - \frac{P\{Y(1,s)=1\}}{P\{Y(0)=1\}},\ \] 
where $P\{Y(0)=1\}$ is the probability of outcome if the whole cohort were assigned to receive placebo. We propose to define a {\it controlled VE CoP} as an immunologic biomarker with $\mathit{CVE}(s)$ non-decreasing in $s$ with $\mathit{CVE}(s) < \mathit{CVE}(s')$ for at least some $s < s'$. 
Because $\mathit{CVE}(s)$ depends on $s$ entirely through $r_C(s)$, a controlled risk CoP and a controlled VE CoP are actually equivalent, and the distinction for applications is whether an unvaccinated/placebo arm is available.  Where available, a controlled VE CoP has the preferred interpretation for most applications such as predicting VE in new settings (bridging). 
  Both types of causal CoPs may either be an `absolute CoP' or a `relative CoP' in the Plotkin nomenclature.\cite{Plotkin2008,Plotkin2010,PlotkinGilbert2018} 

\section{Estimation and testing of controlled risk and control vaccine efficacy}

Since they are causal parameters, controlled risk CoPs and controlled VE CoPs are better fits to the 
provisional-approval goalpost that an immunologic biomarker is reasonably likely to predict vaccine efficacy \cite{FlemingPowers2012} than a CoR association parameter.  We consider estimation of $r_C(s)$, where the estimator $\widehat r_C(s)$ is also used in the estimator of $\mathit{CVE}(s)$.  
Consider the marginalized risk 
\begin{eqnarray}
 r_M(s) = E\left[r(s,X)\right], \label{eq: OVR} 
\end{eqnarray}

\noindent where 
$r(s,x)=P(Y=1\,|\,S=s,A=1,X=x)$ and the outer expectation is over the marginal distribution of $X$ in the study population. The value $r_M(s)$ can be estimated from the observed data without untestable assumptions, and thus $r_M(s)$ is a CoR association parameter.
This parameter averages the biomarker-conditional risk over  
the distribution of $X$ (i.e., direct standardization or g-computation).
If $r_C(s)>r_C(s')$ for $s < s'$, then by definition the risk decreases if the antibody biomarker is increased from $s$ to $s'$.  In contrast, this implication does not necessarily follow if 
$r_M(s) > r_M(s')$ for $s < s'$, because an unmeasured confounder not included in $X$ could create a reversal wherein $r_M(s) > r_M(s')$ even though $r_C(s) < r_C(s')$.

\subsection{Identifiability assumptions}

At biomarker level $s$, the controlled and marginalized risks 
coincide, i.e.,
\begin{eqnarray}
r_C(s)=r_M(s)\, \label{eq: franco}
\end{eqnarray}

\noindent provided $Y(1,s)$ and $S$ are independent given $X$ (no unmeasured confounding), and $P(S=s\,|\,A=1,X)>0$ almost surely (positivity). In other words, identification of the controlled risk curve at biomarker level $s$ requires that a rich enough set of covariates be available so that deconfounding of the relationship between endpoint $Y$ and biomarker $S$ is possible in the population of vaccine recipients, and that $s$ be an observable biomarker level within each subpopulation of vaccine recipients defined by values of $X$.

\subsection{Estimation of the controlled risk curve via regression} 

Various approaches can be used to estimate 
$r_M(s)$, many of which are based on positing a model for $r(s,x)$,
estimating the unknown parameters of this model, and obtaining predicted values $\widehat{r}(s,X_i)$
for each vaccine recipient $i$ with $S_i$ measured.  For example, an inverse-probability-weighted complete-case (IPW-CC) version of this approach estimates $r_M(s)$ by
\comment{
\begin{eqnarray}
\widehat{r}_M(s) = \frac{\sum_{i=i}^n \alpha_i(s)\widehat{r}(s,X_i) }{\sum_{i=i}^n \alpha_i(s)}\mbox{\ \ with\ \ }\alpha_i(s)=\frac{\widehat f(s|X_i)}{\widehat \pi(X_i,Y_i)}\ , \label{eq: obsriskest}
\end{eqnarray}

\noindent where $n$ is the number of vaccine recipients with $M=1$, $\widehat f(s|x)$ is an estimate of the conditional density or probability mass function $f(s|x)$ of $S$ given $A=1$ and $X=x$, and $\widehat \pi(x,y)$ is an estimate of the probability $\pi(x,y) = P(M=1\,|\,Y=y,A=1,X=x)$ of having data on $S$. 
For unbiased estimation, the method for estimating the parameters of the model for $r(s,x)$ must account for the estimated sampling probabilities $\widehat \pi(X_i,Y_i)$ (e.g., \cite{BreslowHolubkov1997,Prentice1986,Barlowetal1999}).  Moreover, the estimator
$\widehat f$ needs to be consistent for $f$, which also requires using weights $1\slash \widehat \pi(X_i,Y_i)$ in the employed conditional density estimation technique.  Moreover, the estimator $\widehat \pi$ needs to be consistent for $\pi$, which often readily holds because the investigator controls which participants are sampled for measurement of $S$.
}
\begin{eqnarray}
\widehat{r}_M(s) = \frac{\sum_{i=i}^n \widehat{r}(s,X_i)\slash \widehat \pi(X_i,Y_i) }{\sum_{i=i}^n 1\slash \widehat \pi(X_i,Y_i)}\ , \label{eq: obsriskest}
\end{eqnarray}

\noindent where $n$ is the number of vaccine recipients with $M=1$ and $\widehat \pi(x,y)$ is an estimate of the probability $\pi(x,y) = P(M=1\,|\,Y=y,A=1,X=x)$ of having data on $S$. 
For unbiased estimation, the method for estimating the parameters of the model for $r(s,x)$ must account for the estimated sampling probabilities $\widehat \pi(X_i,Y_i)$ (e.g., \cite{BreslowHolubkov1997,Prentice1986,Barlowetal1999}).  In addition, the estimator $\widehat \pi$ needs to be consistent for $\pi$, which often readily holds because the investigator controls which participants are sampled for measurement of $S$.

In a setting where only a final binary outcome $Y$ is registered, 
a common approach employs a generalized linear model
or scaled logit model \cite{Dunningetal2015} to estimate $r(s,x)$.
In a survival analysis framework that accounts for loss to follow-up, the outcome may be $Y=I(T \le t)$ for some fixed $t$, with $T$ the time from immunologic biomarker measurement until the clinical endpoint, and 
$r(s,x)$ is modeled, e.g., using a proportional hazards model. Alternatively, 
flexible nonparametric or machine learning-based regression approaches could be used. \cite{westling2020causal,WestlingCarone2020,PriceGilbertvanderLaan2018} 
We aim to provide a reasonable sensitivity analysis framework that can be employed for almost any regression approach used for CoR analysis, including recently proposed approaches.\cite{son2020fast,Sachsetal2020}

\subsection{Conservative bounded estimation of the controlled risk ratio and curve via sensitivity analysis}
\label{sens}

As CoR analysis is based on observational data --- the biomarker value is not randomly assigned ---
a central concern is that unmeasured or uncontrolled confounding of the association between $S$ and $Y$ could render $r_M(s) \neq r_C(s)$, biasing 
estimates of the controlled risk curve $r_C(s)$ and of controlled risk ratios of interest 
$$RR_C(s_1,s_2) = r_C(s_2)/r_C(s_1)\ .$$

Sensitivity analysis is useful to evaluate how strong unmeasured confounding would have to be to explain away an inferred causal association, i.e., inference on $r_M(s)$ indicates an association yet 
$r_C(s)$ is flat.   
As recommended by VanderWeele and Ding (2017), we report the E-value as a summary measure of the evidence of causality.
The E-value is the minimum strength of association, on the risk ratio scale, that an unmeasured confounder would need to have with both the exposure ($S$) and the outcome ($Y$)  to fully explain away a specific observed exposure–outcome association, conditional on the measured covariates.\cite{vanderWeeleDing2017,vanderWeeleMathur2020}  
 Here, ``explained away" means that under the E-value-level of unmeasured confounding the causal effect would be nullified (controlled risk ratio $RR_C(s_1,s_2) = 1$). 
 Because  
$RR_C(s_1,s_2)=(1-CVE(s_2))\slash (1-CVE(s_1)),$ evidence for $RR_C(s_1,s_2) < 1$ is equivalently evidence for $CVE(s_1) < CVE(s_2)$.  Thus in a placebo-controlled trial $RR_C(s_1,s_2)$ can be interpreted as the multiplicative degree of superior vaccine efficacy caused by marker level $s_2$ compared to caused by marker level $s_1$. 

Consider two exposure/marker-level subgroups of interest $S=s_1$ and $S=s_2$ with $s_1 < s_2$, with $\widehat{RR}_M(s_1,s_2) = \widehat{r}_M(s_2)\slash \widehat{r}_M(s_1) < 1$. The E-value for the point estimate $\widehat{RR}_M(s_1,s_2)$ is 
\begin{eqnarray}
e_{pt.est}(s_1,s_2) = \frac{1+\sqrt{1-\widehat{RR}_M(s_1,s_2)}}{\widehat{RR}_M(s_1,s_2)}\ .\label{eq: Evalue}
\end{eqnarray}

\noindent The E-value $e_{UL}(s_1,s_2)$ for the upper 95\% confidence limit $\widehat{UL}(s_1,s_2)$ for $RR_M(s_1,s_2)$ is 
$$min\left\{ 1,\frac{1+\sqrt{1-\widehat{UL}(s_1,s_2)}}{\widehat{UL}(s_1,s_2)} \right\}\ . $$

\noindent 
These two E-values quantify confidence in an immunologic biomarker as a controlled risk CoP and as a controlled VE CoP if there is a placebo arm, with E-values near one suggesting weak support and evidence increasing with greater E-values.  

\comment{
To illustrate the interpretation of an E-value, suppose $S$ is binary and
$\widehat{RR}_M(0,1) = \widehat{r}_M(1) \slash \widehat{r}_M(0) = 0.40$ with 95\% confidence interval (CI) $(0.14, 0.78)$.  An E-value $e(0,1)=4.4$ means that the marginalized risk ratio $RR_M(0,1)$ at point estimate 0.40 could be explained away by an unmeasured confounder 
associated with both the exposure and the outcome by
a marginalized risk ratio of 4.4-fold each, after accounting for the 
measured confounders $X$, but that weaker confounding could
not do so.  Here ``explained away" means a nullified causal effect, i.e., point estimate of $RR_C(0,1)=1.0$.  

The E-value $e_{UL}(s_1,s_2)$ for the upper limit $\widehat{UL}(s_1,s_2)$ of the 95\% CI for $\widehat{RR}_M(s_1,s_2)$ is also reported, computed as 1 if $\widehat{UL}(s_1,s_2)\ge 1$ and, otherwise, as
$$\frac{1+\sqrt{1-\widehat{UL}(s_1,s_2)}}{\widehat{UL}(s_1,s_2)}\ , $$

\noindent which in the example equals $e_{UL}(0,1)=1.88$.  
}

It is also useful to provide conservative estimates of controlled risk ratios and of the controlled risk curve, accounting for unmeasured confounding.  
We approach these tasks based on the sensitivity/bias analysis approach of Ding and VanderWeele (2016).\cite{DingvanderWeele2016} 
Define two context-specific 
sensitivity parameters for any given $s_1 < s_2$: 
$RR_{UD}(s_1,s_2)$
is the maximum risk ratio for the outcome $Y$ comparing any two categories of the unmeasured confounder $U$, within either exposure group $S=s_1$ or $S=s_2$, conditional on the
vector $X$ of observed covariates; and $RR_{EU}(s_1,s_2)$ is the maximum risk
ratio for any specific level of the unmeasured confounder $U$ comparing individuals
with $S=s_1$ to those with $S=s_2$,
with adjustment already made for the measured covariate vector $X$
(where $RR_{UD}(s_1,s_2) \ge 1$ and $RR_{EU}(s_1,s_2) \ge 1$).
Thus, $RR_{UD}(s_1,s_2)$ quantifies the importance of the unmeasured confounder $U$ for the outcome, and $RR_{EU}(s_1,s_2)$ quantifies how imbalanced the exposure/marker subgroups $S=s_1$ and $S=s_2$ are in the
unmeasured confounder $U$.    
We suppose that $RR_M(s_1,s_2) < 1$ for the values $s_1 < s_2$ used in a data analysis --- this is the case of interest for immune correlates assessment.

Define the bias factor 
$$B(s_1,s_2) = \frac{RR_{UD}(s_1,s_2) RR_{EU}(s_1,s_2)}{RR_{UD}(s_1,s_2) + RR_{EU}(s_1,s_2) - 1}\ $$

for $s_1 \le s_2$, and define $RR^U_M(s_1,s_2)$ the same way as
$RR_M(s_1,s_2)$, except marginalizing over the joint distribution of $X$ and $U$. Then, $RR^U_M(s_1,s_2)\leq RR_M(s_1,s_2)\times B(s_1,s_2)$, where $RR^U_M(s_1,s_2) = E\{r(s_2,X^*)\}/ 
E\{r(s_1,X^*)\}$ with $X^* = (X,U)$ and $r$ conditional risk defined near equation  (\ref{eq: OVR}).\cite{DingvanderWeele2016}
Translating this result into our context, under the positivity assumption $RR^U_M(s_1,s_2) = RR_C(s_1,s_2)$, so that 
\begin{eqnarray}
RR_C(s_1,s_2) \le RR_M(s_1,s_2)\times B(s_1,s_2)\ . \label{eq: Dingvan2016result}
\end{eqnarray}

\noindent It follows that a conservative (upper bound) estimate of $RR_C(s_1,s_2)$ is $\widehat{RR}_M(s_1,s_2)\times B(s_1,s_2)$, and a conservative 95\% CI is obtained by multiplying each confidence limit for
$RR_M(s_1,s_2)$ by $B(s_1,s_2)$.  These estimates 
account for the presumed-maximum plausible amount of deviation from the no unmeasured confounders assumption specified by user-supplied values of $RR_{UD}(s_1,s_2)$ and $RR_{EU}(s_1,s_2)$.
 An advantage of this approach is that the bound (\ref{eq: Dingvan2016result}) holds without making any assumption about the confounder vector $X$ or the unmeasured confounder $U$.\nocite{DingvanderWeele2016}  

To provide conservative inference for $r_C(s)$, we next select a central value $s^{cent}$ of $S$ such that $\widehat{r}_M(s^{cent})$ matches the observed overall risk, 
$\widehat P(Y=1|A=1)$.
This value is a `central' marker value at which the observed marginalized risk equals the observed overall risk.  Next, we `anchor' the analysis by assuming
$r_C(s^{cent}) = r_M(s^{cent}),$ where picking the central value $s^{cent}$ makes this plausible to be at least approximately true.
Under this assumption, the bound (\ref{eq: Dingvan2016result}) implies the bounds 
\begin{eqnarray}
r_C(s) &\le &r_M(s) B(s^{cent},s) \hspace{.1in} \hbox{\ if} \hspace{.1in} s\ge s^{cent} \label{eq: biasform1} \\
r_C(s) &\ge &r_M(s) \frac{1}{B(s,s^{cent})} \hspace{.1in} \hbox{\ if} \hspace{.1in} s < s^{cent}. \label{eq: biasform2}
\end{eqnarray}

\noindent Therefore, after specifying $B(s^{cent},s)$ and $B(s^{cent},s)$ for all $s$, we conservatively estimate $r_c(s)$ by plugging $\widehat r_M(s)$ into the formulas (\ref{eq: biasform1}) and (\ref{eq: biasform2}). 
Because $B(s_1,s_2)$ is always greater than one for $s_1 < s_2$,
formula (\ref{eq: biasform1}) pulls the observed risk $\widehat{r}_M(s)$ upwards for subgroups with high biomarker values, and 
formula (\ref{eq: biasform2}) pulls the observed risk $\widehat{r}_M(s)$ downwards for subgroups
with low biomarker values. This makes the estimate of the controlled risk curve flatter, closer to the null curve, as desired for a sensitivity/robustness analysis.

To specify $B(s_1,s_2)$, we note that it should have greater magnitude for a greater distance of $s_1$ from $s_2$, as determined by specifying 
$RR_{UD}(s_1,s_2)$ and $RR_{EU}(s_1,s_2)$ increasing with
$s_2 - s_1$ (for $s_1 \le s_2$).  
We consider one specific approach, which sets
$RR_{UD}(s_1,s_2) = RR_{EU}(s_1,s_2)$ to the common value $RR_U(s_1,s_2)$ that is specified log-linearly: $\log RR_U(s_1,s_2)$ \newline $= \gamma (s_2 - s_1)$ for $s_1 \le s_2$.  
Then, for a user-selected pair of values $s_1=s^{fix}_1$ and $s_2=s^{fix}_2$ with $s^{fix}_1 < s^{fix}_2$, 
we set a sensitivity parameter
$RR_U(s^{fix}_1,s^{fix}_2)$ to some value above one.  
It follows that   
$$\log RR_U(s_1,s_2) = \left(\frac{s_2 - s_1}{s^{fix}_2 - s^{fix}_1}\right) \log RR_U(s^{fix}_1,s^{fix}_2), \hspace{.2in} s_1 \le s_2.$$

\begin{figure}
\begin{center}
\includegraphics[width=\textwidth,angle=0]{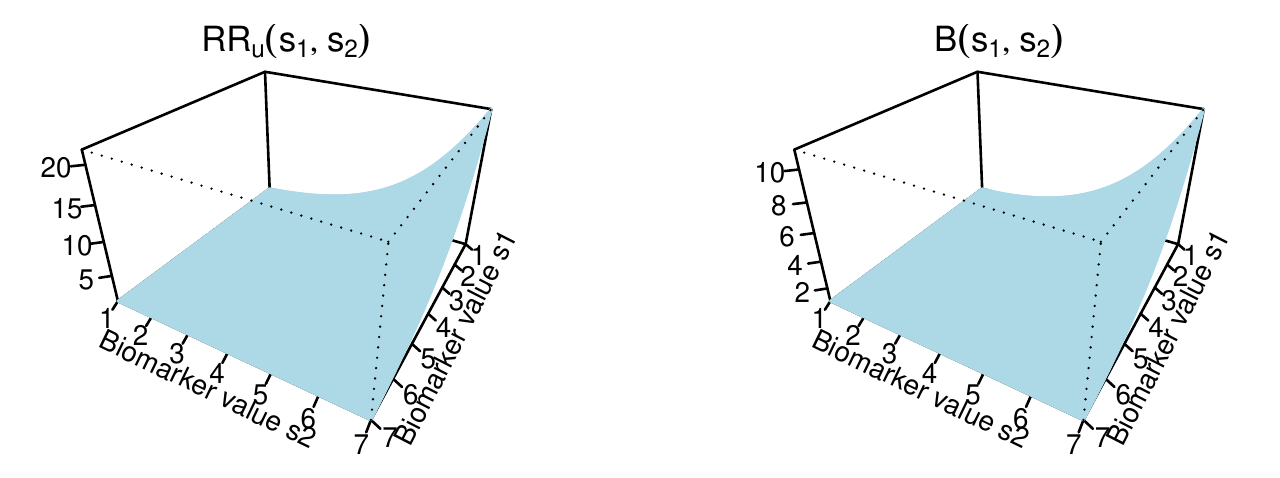}
\end{center}
\caption{$RR_U(s_1,s_2)$ and $B(s_1,s_2)$ surfaces for $s_1 \le s_2$ with user-supplied sensitivity parameter $RR_U(s^{fix}_1,s^{fix}_2) = 4$ with $s^{fix}_1=s_{med}$ the median of $S$ and $s^{fix}_2=s_{0.95}$ the 95th percentile
of $S$ (the specified degree of unmeasured confounding) where $RR_U(s_1,s_2) = RR_{UD}(s_1,s_2) = RR_{EU}(s_1,s_2)$}\label{RRuplot}

\end{figure}

\noindent Figure \ref{RRuplot} illustrates the $RR_U(s_1,s_2)$ and $B(s_1,s_2)$ surfaces.


\subsection{Use of marginalized risk and generality of the approach}

Our approach focuses on marginalized covariate-adjusted association parameters.  This choice provides a simple conversion of the association parameters to causal parameters through the equation $r_M(s)=r_C(s)$. 
Moreover, 
 a very common approach to covariate-adjusted CoR analysis consists of fitting a two-phase sampling regression 
 (e.g., logistic, Cox) 
 model to the immunologic biomarker and observed confounders, and subsequently reporting estimates of the association parameter (e.g., odds ratio, hazard ratio) corresponding to the immunologic biomarker covariate.
Because odds ratios and hazard ratios are not collapsible, 
the conditional odds ratio and conditional hazard ratio do not in general equal the marginalized odds ratio and marginalized hazard ratio, respectively. \cite{RobinsGreenland1992,loux2017comparison}  The dependency of the conditional odds/hazard ratio on the set of confounders conditioned upon may make these parameters less useful for generalizability of inferences than the marginalized parameters. 
 Yet, the same approach applies for alternative marginalized parameters, e.g. the marginalized odds ratio  $$\frac{r_M(s_2)/\{1-r_M(s_2)\}}{r_M(s_1)/\{1-r_M(s_1)\}}\ .$$  The same specification of 
$RR_{UD}(s_1,s_2)$ and $RR_{EU}(s_1,s_2)$ can be used, as the outcome in vaccine efficacy trials is generally rare.  \cite{DingvanderWeele2016,vanderWeeleDing2017}  

\subsection{Estimation of controlled vaccine efficacy with sensitivity analysis}

  As evidence for a controlled VE CoP should be robust to potential bias from unmeasured confounding, we propose reporting conservative estimates and confidence intervals for $\mathit{CVE}(s)$, where `conservative' means including margin for potential unmeasured confounding that makes the estimated curve $\mathit{CVE}(s)$ flatter.  To do this, we write
$\mathit{CVE}(s)= 1 - r_C(s) \slash E[P(Y=1|A=0,X)]$, and estimate $r_C(s)$ based on the equations
(\ref{eq: biasform1}) and (\ref{eq: biasform2}). 
The denominator (placebo arm marginalized risk) equals $P\{Y(0)=1\}$ based on the randomization; there is thus no concern about unmeasured confounding.  While this denominator may be estimated validly ignoring $X$, we favor using an estimation approach compatible with whatever approach was used to estimate $r_M(s)$.
For most regression approaches to estimation of $r_M(s)$ the bootstrap can be used to obtain valid 95\% confidence intervals for $\mathit{CVE}(s)$, implemented in the same manner as for inference about $r_M(s)$ and $r_C(s)$ except that the placebo arm must also be bootstrapped.

If the study cohort is naive to the relevant pathogen (e.g., SARS-CoV-2 VE trial primary cohorts), then the marker $S$ has no variability in the placebo arm [all values are `negative,' below the assay lower limit of detection (LLOD)].  Advantageously, in this setting $\mathit{CVE}(s)$ has a special connection to the natural effects mediation literature,\nocite{Cowlingetal2019} where 
$\mathit{CVE}(s < LLOD)$ is the natural direct effect, and 100\% of vaccine efficacy is 
mediated through $S$ if and only if $\mathit{CVE}(s < LLOD) = 0$. Thus inference on $\mathit{CVE}(s < LLOD)$ evaluates full mediation.

\section{Analysis of two dengue vaccine efficacy trials}\label{sec:CYD14CYD15}

In the CYD14 (NCT01373281)\cite{Capedingetal2014} and CYD15 (NCT01374516)\cite{Villaretal2015} trials, participants were randomized  2:1 to the CYD-TDV dengue vaccine vs. placebo, with immunizations at Months 0, 6, and 12.  The primary analyses assessed vaccine efficacy (VE) against symptomatic, virologically confirmed dengue (VCD) occurring at least 28 days after the third immunization through to the Month 25 visit. Based on a proportional hazards model, estimated VE was 56.5\% (95\% CI 43.8--66.4) in CYD14 and 64.7\% (95\% CI 58.7--69.8) in CYD15.  Sanofi Pasteur conducted the CYD14 and CYD15 trials and provided access to the data.

Month 13 (M13) neutralizing antibody (nAb) titers to each serotype were measured through case-cohort Bernoulli random sampling of all randomized participants at enrollment and from all participants who experienced the VCD endpoint after M13 and by Month 25 (cases).  A participant's average M13 log10-transformed geometric mean titer across serotypes (``M13 average titer") has been studied as a CoR of VCD;\cite{Vigne2017,Moodieetal2018} we study this biomarker as a controlled risk CoP and as a controlled VE CoP.

For estimation of $r(s,x)$, we apply a method used in Moodie et al.,\cite{Moodieetal2018} 
Cox partial likelihood regression \cite{Prentice1986} that accounted for the case-cohort sampling design.
The baseline covariates $X$ accounted for are protocol-specified age categories, sex, and country.  Participants without VCD through the M13 visit and with
M13 average titer measured are included in analyses.

We first investigate how the marginalized risk of VCD compares between vaccine recipients with highest versus lowest tertile values of M13 average titer, coded $s=1$ and $s=0$.
The tasks of CoR analysis are to estimate $r_M(s=1)$, $r_M(s=0)$, and 
$RR_M(0,1) = r_M(s=1) \slash r_M(s=0)$.  Given the sampling design, $\widehat \pi(X_i,Y_i)$ is 1.0 for all cases $Y_i=1$ and is 0.195 (0.096) for 
non-cases $Y_i=0$ sampled into the CYD14 (CYD15) subcohort.  We estimate each $r_M(s)$ based on equation (\ref{eq: obsriskest}).
 We use the bootstrap to obtain 95\% pointwise confidence intervals for each marginalized risk and marginalized risk ratio, which directly provide confidence intervals for each controlled risk and controlled risk ratio assuming no unmeasured confounding.  Under the E-value formulas they also provide results assuming a certain amount of unmeasured confounding.
The results in Table 1 show  
E-values for $RR_C(0,1)$ much larger than 1.0 with 95\% CIs lying below 1.0 for both trials, supporting a controlled risk CoP robust to unmeasured confounding. 
For instance, the CYD15 results with 95\% CI for $RR_C(0,1)$ 0.04--0.20 building in margin for unmeasured confounding can be interpreted as marker level in the third vs. first tertile causing at least 5 times (5 = 1/0.20) greater vaccine efficacy, accounting for uncertainty both due to sampling variability and to unmeasured confounding. 

\begin{center}
\begingroup
\renewcommand{\arraystretch}{2} 
\begin{table}[H] \centering
\doublespacing
\caption{Analysis of M13 average titer (upper vs. lower tertile) as a CoR and a controlled risk CoP: CYD14 and CYD15 dengue vaccine efficacy trials}
\label{table1}
\singlespacing
\begin{tabular}{lcccccc} \hline \hline
 & \multicolumn{2}{c}{marginalized risk} & \multicolumn{2}{c}{controlled risk} & & \\ 
 & \multicolumn{2}{c}{ratio $RR_M(0,1)$} & \multicolumn{2}{c}{ratio $RR_C(0,1)^1$} & \multicolumn{2}{c}{e(0,1)$^2$} \\ 
Trial  & Point Est. & 95\% CI  &    Point Est. & 95\% CI  &   Point Est.  & 95\% CI UL \\ \hline
CYD14 & 0.16&0.08--0.29&0.38&0.18--0.66&11.6&6.4
   \\ 
CYD15 & 0.05&0.02--0.09&0.10&0.04--0.20&43.7&22.0
   \\ \hline \hline
\end{tabular}
\newline
\doublespacing
\noindent $^1$Conservative (upper bound) estimate assuming unmeasured confounding at level $RR_{UD}(0,1)=RR_{EU}(0,1) = 4$ and thus $B(0,1)=16/7$. \newline
\noindent $^2$E-values are computed for upper tertile $s=1$ vs. lower tertile $s=0$ biomarker subgroups after controlling for age, sex, and country; UL = upper limit.
\end{table}
\endgroup
\end{center}

\comment{
$RR(0,1)$ are $\widehat{RR}(1,0) = xx (95\% CI xx-xx)$ for CYD14 and 
$\widehat{RR}(0,1) = xx (95\% CI xx-xx)$ for CYD15.  For CYD14, the E-value $e_{RR}(0,1)$ is xx, with E-value $e_{UL}(0,1)$ for the upper limit equal to xx.  For CYD15, the E-value $e_{RR}(0,1)$ is xx, with the E-value $e_{UL}(0,1)$ for the upper limit equal to xx.  Next, we set $RR_{UD}(0,1)=RR_{EU}(0,1)=2$, such that $B(0,1)=$.  The resulting conservative estimates of the controlled risk ratio are $\widehat{RR_C}(0,1) = xx (95\% CI xx-xx)$ for CYD14 and 
$\widehat{RR_C}(0,1) = xx (95\% CI xx-xx)$ for CYD15 (PENDING: Just multiply by $B(0,1)$).  
}

Next we repeat the analysis treating $S$ as a quantitative variable, where $r(s,x)$ is again estimated by two-phase Cox partial likelihood regression and now $RR_M(s_1,s_2)$ is the marginalized risk ratio between $s_1$ and $s_2$. Let $s_1$ and $s_2$ be the 15$^{th}$ and 85$^{th}$ percentile of M13 average titer.  
The results for $RR_M(s_1,s_2)$ are $\widehat{RR}_M(s_1,s_2) = 0.20%
$ $(95\%$ CI 0.12--0.31%
) for CYD14 and  $\widehat{RR}_M(s_1,s_2) = 0.13%
$ $(95\%$ CI 0.09--0.18%
) for CYD15.  For CYD14, the E-value $e_{pt.est}(s_1,s_2)$ is 9.3%
, with E-value $e_{UL}(s_1,s_2)$ for the upper limit equal to 5.9%
.  For CYD15, the E-value $e_{pt.est}(s_1,s_2)$ is 14.6%
, with E-value $e_{UL}(s_1,s_2)$ for the upper limit equal to 10.5%
.  Next, we set $RR_{UD}(s_1,s_2)=RR_{EU}(s_1,s_2)=4$, such that $B(s_1,s_2)=16/7$.  The resulting upper bound estimates of the controlled risk ratio are $\widehat{RR_C}(s_1,s_2) = 0.47%
$ $(95\%$ CI 0.28--0.71%
) for CYD14 and $\widehat{RR_C}(s_1,s_2) = 0.30%
$ $(95\%$ CI 0.19--0.42%
) for CYD15.  Figure 2 shows 
 the sensitivity analysis described in Section 
\ref{sens}.  
 After building in the margin for unmeasured confounding, for CYD4 estimated VCD risk decreases from 0.032 at 
low M13 average titer value 1:36 to 0.01 at high titer value 1:1200; for CYD15 these results are 0.033 and 0.008, respectively. 

Figure \ref{CoPveryhighVE_Fig3} shows results for
the 
$\mathit{CVE}(s)$ curve,
 where $E[P(Y=1|A=0,X)]$ was estimated with a standard Cox model 
with point estimate the average of the fitted values $\widehat E[P(Y_i=1|A_i=0,X_i)]$ across all placebo recipients at-risk for VCD at the M13 visit. 
The results show that after accounting for potential unmeasured confounding the point estimate of $\mathit{CVE}(s)$ in CYD14 monotonically increases from 29.6\% (95\% CI  3.5--45.9)
 at low M13 average titer value of 36 to 78.5\% (95\% CI 67.9--86.8)
 at high value 1200. The degree of increase is greater for CYD15: from  17.4\% (95\% CI -14.4-- 36.5)
 to 84.5\% (95\% CI 79.6--89.1)
. This supports a robust controlled VE CoP. 
The steeper increase and potentially superior CoP for CYD15 may be due to the older ages (9--16 vs. 2--14) and corresponding larger fraction of previously-infected vaccine recipients.  
The $CVE(s)$ curve is above zero at lowest average titer, suggesting the CoP imperfectly mediates VE. This could be due to a variety of reasons including that the marker does not fully capture immune response quality.

For validity the analyses require positivity; Price et al.\cite{PriceGilbertvanderLaan2018} showed that M13 average titer in vaccine recipients varied over its whole range at each covariate level of $X$ in CYD14 and in CYD15, supporting this assumption.
In conclusion, the evidence for M13 nAb titer as a controlled risk CoP and as a controlled VE CoP is robust to no unmeasured confounding.  As some vaccine recipients with high titer experienced VCD, the CoP is a relative CoP, not an absolute CoP.
\section{Discussion}

In virtually all immune correlates analyses of vaccine efficacy trials or prospective cohort studies, immunologic biomarkers are studied as correlates of risk in vaccine recipients.  Given the goal to establish a biomarker as a correlate of protection/surrogate endpoint, it is generally of interest to evaluate the extent to which the correlates of risk results --- which are based on associational parameters --- can be interpreted in terms of causal effects, making them constitute a more reliable basis for decision-making for various 
vaccine application questions.  We proposed a general approach to augment correlates of risk analysis with a conservative analysis of the controlled risk curve $r_C(s)$, and also of the controlled vaccine efficacy curve $\mathit{CVE}(s)$ if there was a randomized placebo arm.  The approach is general because it allows any frequentist method for estimation of the underlying conditional risk curve. 
 It is conservative by providing estimates and confidence intervals that assume a specified amount of unmeasured confounding that make it more difficult to conclude a controlled risk CoP (equivalently a controlled vaccine efficacy CoP if there is a placebo arm). For the controlled risk ratio and its relative controlled vaccine efficacy counterpart, the conservative estimation and inference is achieved by reporting E-values for the point estimate and upper 95\% confidence limit, whereas for the controlled risk and controlled vaccine efficacy curves it is quantified through an interpretable parametrization of the bias function that makes the controlled risk estimate flatter/closer to the null. 

Aligned with VanderWeele and Ding (2017) \cite{vanderWeeleDing2017}, we do not think a particular E-value magnitude indicating a truly robust result should be pre-specified, as the interpretation of the E-value depends on the problem context, including the study endpoint, the vaccine, the set of potential confounders that are adjusted for, and the plausible magnitude of unmeasured confounders.  
For example, if there is extensive knowledge about the endpoint prognostic factors and a rich set of baseline potential confounders are collected, then the bar for the magnitude of the E-value to be convincing will be smaller.
The assessment of controlled risk / vaccine efficacy CoPs should include the study of the strength of observed confounding as context for interpreting potential unmeasured confounding.
For SARS-CoV-2 vaccine studies, this would include studying the association of age, co-morbidities, and race/ethnicity with infection or disease outcomes.
It is useful for the sensitivity analysis to include a table reporting associations of potential confounders with outcomes and with immunologic biomarkers. 



Our sensitivity analysis through E-values and the bias function only addresses unmeasured confounding; violation of the positivity assumption, selection bias, and missing data could also make the controlled risk CoP analysis results misleading.  For prospective cohort studies with investigator control of which blood samples are selected for biomarker measurement, selection bias and missing data should not be a problem, unless a large percentage of participants have missing blood samples at the key biomarker sampling time point.  
In our approach, the positivity assumption is required to convert the estimation of an associational parameter into estimation of a causal parameter, and so, we recommend the approach only be applied if diagnostics support that the antibody biomarker in vaccine recipients tends to vary over its full range within each level of the potential confounders that are adjusted for.  Alternatively, specialized methods designed to be more robust to positivity violations (e.g., collaborative targeted minimum loss-based estimation \cite{van2010collaborative}) could be incorporated into our proposed framework.

Additionally, a basic premise of the proposed approach is that it is conceivable to assign every vaccine recipient to have an immunologic biomarker set to a given value $s$.
Diagnostics may be helpful to examine this premise. For example, if the oldest and/or immunocompromised participants tend to have low immune responses, it may be difficult to conceive of assigning their biomarker to the highest levels of $S$.
One way to address this problem would be to base CoP evaluation on the stochastic interventional risk curve defined in Hejazi et al. \cite{Hejazietal2020} instead of the controlled risk approach outlined here.  This approach sets each vaccinated recipient's biomarker based on a random draw from a specified distribution --- for example, from the observed distribution with a specified shift upwards for some or all participants --- which in some settings may be more plausible. 

While we focused on controlled vaccine efficacy based on the controlled risk curve $P\{Y(1,s)=1\}$ versus $s$, it may  also be interesting to study 
$r^+_C(s)=P\{Y(1,s+)=1\}$ versus $s$, where $Y(1,s+)$ denotes risk under assignment to vaccine and 
to the biomarker $S \ge s$.  In the dengue application, we studied this curve for two tertile cut-points. If the biomarker is continuous, similar regression methods can be used for estimation and inference on $r^+_C(s)$ over the entire span of $s$ values.  One advantage of studying $r^+_C(s)$ is that regulators have historically defined immunologic surrogate endpoints for approval decisions in terms of thresholds \cite{Plotkin2008,Plotkin2010}. 
However, since assignment to a level $S\geq s$ does not uniquely define a hypothetical intervention, the interpretation of the resulting controlled risk $r^+_C(s)$ will typically hinge on the degree to which each level of $S$ above $s$ is represented in the study sample --- this can result in less interpretable scientific findings.

In our application, we focused on two-phase Cox regression, a commonly used method, as the basis of the analysis of the marginalized risk curve $r_M(s)$.  Yet, given the critical need to control for confounding and minimize systematic bias, it may be appealing to alternatively incorporate into the analysis nonparametric doubly-robust methods for estimating $r_M(s)$ under minimal assumptions, along the lines of
Westling et al. (2020) \cite{westling2020causal}.  

 Among surrogate endpoint evaluation approaches that have been considered, the principal stratification approach may most closely resemble the controlled vaccine efficacy approach, because it estimates a `vaccine efficacy curve' across principal strata defined by the level $s$ of the immunologic biomarker if assigned vaccine, and the outcome of the analysis looks similar.  However, the two approaches address different scientific objectives, with principal stratification assessing vaccine efficacy across `natural' subgroups without assignment/intervention on the marker, whereas controlled effects assign the marker level. Thus an advantage of principal stratification is obviating the need to conceive assignment of a biomarker that was not randomly assigned; yet this implies the limitation of not fully addressing causal mediation of vaccine efficacy (e.g., VanderWeele, 2008).\nocite{vanderWeele2008}  Consequently, while it is more challenging to conceptualize and define controlled vaccine efficacy, if this challenge is overcome then inferences are more relevant for core applications such as bridging -- predicting vaccine efficacy of new vaccines based on the immunologic biomarker.

Lastly, 
the science of the biological assays used to define immunologic biomarkers should be highlighted as 
fundamental to the establishment and utility of a CoP.  Pre-specified validation criteria are typically required for an immunologic biomarker to be accepted as a surrogate endpoint.\cite{FDAGuidance2018}  Post-acceptance, effective use of a surrogate endpoint requires that the biomarker be measured by the same lab that established the CoP, or that a new lab conducting the immunoassay has validated concordance of its assay compared to the original assay conducted by the original lab.  Thus, standardization and validation of the immunoassay used to measure the CoP is a basic requirement for use of a CoP for approving/bridging vaccines.

\bibliography{ref}

\begin{thebibliography}{10}

\bibitem{polack2020safetyshort}
Fernando~P Polack, Stephen~J Thomas, Nicholas Kitchin, Judith Absalon,
  Alejandra Gurtman, Stephen Lockhart, John~L Perez, Gonzalo P{\'e}rez~Marc,
  Edson~D Moreira, Cristiano Zerbini, et~al.
\newblock Safety and efficacy of the {BNT162b2 mRNA Covid-19} vaccine.
\newblock {\em New England Journal of Medicine}, 383(27):2603--2615, 2020.

\bibitem{Badenetal2020short}
Lindsey~R Baden, Hana~M El~Sahly, Brandon Essink, Karen Kotloff, Sharon Frey,
  Rick Novak, David Diemert, Stephen~A Spector, Nadine Rouphael, C~Buddy
  Creech, et~al.
\newblock Efficacy and safety of the {mRNA-1273 SARS-CoV-2} vaccine.
\newblock {\em New England Journal of Medicine}, 2020.

\bibitem{sadoff2021safety}
Jerald Sadoff, Glenda Gray, An~Vandebosch, Vicky C{\'a}rdenas, Georgi Shukarev,
  Beatriz Grinsztejn, Paul~A Goepfert, Carla Truyers, Hein Fennema, Bart
  Spiessens, et~al.
\newblock Safety and efficacy of single-dose ad26. cov2. s vaccine against
  covid-19.
\newblock {\em New England Journal of Medicine}, 2021.

\bibitem{Follmannetal2020}
D~Follmann, J~Fintzi, MP~Fay, HE~Janes, L~Baden, H~El Sahly, TR~Fleming,
  DV~Mehrotra, LN~Carpp, M~Juraska, D~Benkeser, D~Donnell, Y~Fong, S~Han,
  I~Hirsch, Y~Huang, Y~Huang, O~Hyrien, A~Luedtke, M~Carone, M~Nason,
  V~Vandebosch, H~Zhou, I~Cho, E~Gabriel, JG~Kublin, MS~Cohen, L~Corey,
  PB~Gilbert, and KM~Neuzil.
\newblock Assessing durability of vaccine effect following blinded crossover in
  {COVID-19} vaccine efficacy trials.
\newblock {\em medRxiv}, doi: https://doi.org/10.1101/2020.12.14.20248137,
  2020.

\bibitem{FlemingPowers2012}
Thomas~R Fleming and John~H Powers.
\newblock Biomarkers and surrogate endpoints in clinical trials.
\newblock {\em Statistics in Medicine}, 31(25):2973--2984, 2012.

\bibitem{FDASIA}
{US Code of Federal Regulations {FDA} Subpart H} -- {A}ccelerated approval of
  new drugs for serious or life-threatening illnesses.
\newblock {\em 21 CFR}, Secs. 314.500–314.560.

\bibitem{Plotkin2008}
Stanley~A Plotkin.
\newblock Vaccines: Correlates of vaccine-induced immunity.
\newblock {\em Clinical Infectious Diseases}, 47(3):401--409, August 2008.

\bibitem{Plotkin2010}
Stanley~A Plotkin.
\newblock Correlates of protection induced by vaccination.
\newblock {\em Clinical Vaccine Immunology}, 17(7):1055--1065, May 2010.

\bibitem{PlotkinGilbert2018}
S~Plotkin and Peter~B Gilbert.
\newblock Correlates of protection.
\newblock In S~Plotkin, Walter Orenstein, Paul Offit, and Kathryn Edwards,
  editors, {\em Vaccines, Seventh Edition}, pages 35--40. Elsevier Inc., New
  York, 2018.

\bibitem{mason1973yellow}
Richard~A Mason, Nicola~M Tauraso, Richard~O Spertzel, and Robert~K Ginn.
\newblock Yellow fever vaccine: direct challenge of monkeys given graded doses
  of 17d vaccine.
\newblock {\em Applied Microbiology}, 25(4):539--544, 1973.

\bibitem{van2011correlation}
Yvonne Van~Gessel, Christoph~S Klade, Robert Putnak, Alessandra Formica,
  Somporn Krasaesub, Martin Spruth, Bruno Cena, Anchalee Tungtaeng, Montip
  Gettayacamin, and Shailesh Dewasthaly.
\newblock Correlation of protection against japanese encephalitis virus and
  {JE} vaccine {(IXIARO{\textregistered})} induced neutralizing antibody
  titers.
\newblock {\em Vaccine}, 29(35):5925--5931, 2011.

\bibitem{beasley2016first}
David~WC Beasley, Trevor~L Brasel, and Jason~E Comer.
\newblock First vaccine approval under the {FDA Animal Rule}.
\newblock {\em NPJ Vaccines}, 1(1):1--3, 2016.

\bibitem{hobson1972role}
DRAA Hobson, RL~Curry, AS~Beare, and A~Ward-Gardner.
\newblock The role of serum haemagglutination-inhibiting antibody in protection
  against challenge infection with influenza a2 and b viruses.
\newblock {\em Epidemiology \& Infection}, 70(4):767--777, 1972.

\bibitem{chen2016single}
Wilbur~H Chen, Mitchell~B Cohen, Beth~D Kirkpatrick, Rebecca~C Brady, David
  Galloway, Marc Gurwith, Robert~H Hall, Robert~A Kessler, Michael Lock,
  Douglas Haney, et~al.
\newblock Single-dose live oral cholera vaccine {CVD 103-HgR} protects against
  human experimental infection with {Vibrio} cholerae o1 {El Tor}.
\newblock {\em Clinical Infectious Diseases}, 62(11):1329--1335, 2016.

\bibitem{Qinetal2007}
L~Qin, Peter~B Gilbert, L~Corey, J~McElrath, and SG~Self.
\newblock A framework for assessing immunological correlates of protection in
  vaccine trials.
\newblock {\em The Journal of Infectious Diseases}, 196:1304--1312, 2007.

\bibitem{Prentice1989}
RL~Prentice.
\newblock Surrogate endpoints in clinical trials: definition and operational
  criteria.
\newblock {\em Statistics in Medicine}, 8:431--440, 1989.

\bibitem{Freedman1992}
LS~Freedman, BI~Graubard, and A~Schatzkin.
\newblock Statistical validation of intermediate endpoints for chronic
  diseases.
\newblock {\em Statistics in Medicine}, 11:167--178, 1992.

\bibitem{Molenberghsetal2008}
G~Molenberghs, T~Burzykowski, A~Alonso, P~Assam, A~Tilahum, and M~Buyse.
\newblock The meta-analytic framework for the evaluation of surrogate endpoints
  in clinical trials.
\newblock {\em Journal of Statistical Planning and Inference}, 138:432--449,
  2008.

\bibitem{GilbertGabrieletal2014}
Peter~B Gilbert, E~Gabriel, X~Miao, X~Li, S-C Su, and ISF Chan.
\newblock Fold rise in antibody titers by measured by glycoprotein-based
  enzyme-linked immunosorbent assay is an excellent correlate of protection for
  a herpes zoster vaccine, demonstrated via the vaccine efficacy curve.
\newblock {\em The Journal of Infectious Diseases}, 10:1573--1581, 2014.

\bibitem{Moodieetal2018}
Z~Moodie, M~Juraska, Y~Huang, Y~Zhuang, Y~Fong, LN~Carpp, SG~Self,
  L~Chambonneau, R~Small, N~Jackson, F~Noriega, and Peter~B Gilbert.
\newblock Neutralizing antibody correlates analysis of tetravalent dengue
  vaccine efficacy trials in {Asia} and {Latin America}.
\newblock {\em Journal of Infectious Diseases}, 217(5):742--753, 2018.

\bibitem{Cowlingetal2019}
BJ~Cowling, WW~Lim, RA~Perera, VJ~Fang, GM~Leung, JM~Peiris, and
  EJ~Tchetgen~Tchetgen.
\newblock Influenza hemagglutination-inhibition antibody titer as a mediator of
  vaccine-induced protection for influenza b.
\newblock {\em Clinical Infectious Diseases}, 68(10):1713--7, 2019.

\bibitem{Hejazietal2020}
N~Hejazi, MJ~van~der Laan, HE~Janes, PB~Gilbert, and DC~Benkeser.
\newblock Efficient nonparametric inference on the effects of stochastic
  interventions under two-phase sampling, with applications to vaccine efficacy
  trials.
\newblock {\em Biometrics}, https://doi.org/10.1111/biom.13375, 2020.

\bibitem{GabrielDanielsHalloran2016}
Halloran~ME Gabriel~EE, Daniels~MJ.
\newblock Comparing biomarkers as trial level general surrogates.
\newblock {\em Biometrics}, 72:1046--1054, 2016.

\bibitem{Chan2002}
ISF Chan, L~Shu, H~Matthews, C~Chan, R~Vessey, J~Sadoff, and J~Heyse.
\newblock Use of statistical models for evaluating antibody response as a
  correlate of protection against varicella.
\newblock {\em Statistics in Medicine}, 21:3411--3430, 2002.

\bibitem{LiParnesChan2013}
S~Li, M~Parnes, and ISF Chan.
\newblock Determining the cutoff based on a continuous variable to define two
  populations with application to vaccines.
\newblock {\em Journal of Biopharmaceutical Statistics}, 23:662--680, 2013.

\bibitem{mehrotra2020clinical}
Devan~V Mehrotra, Holly~E Janes, Thomas~R Fleming, Paula~W Annunziato,
  Kathleen~M Neuzil, Lindsay~N Carpp, David Benkeser, Elizabeth~R Brown, Marco
  Carone, Iksung Cho, et~al.
\newblock Clinical endpoints for evaluating efficacy in {COVID-19} vaccine
  trials.
\newblock {\em Annals of Internal Medicine}, 2020.

\bibitem{vanderWeele2013}
TJ~VanderWeele.
\newblock Surrogate measures and consistent surrogates.
\newblock {\em Biometrics}, 69:561--568, 2013.
\newblock PMCID: PMC4221255.

\bibitem{FlemingDeMets1996}
TR~Fleming and DL~DeMets.
\newblock Surrogate endpoints in clinical trials: Are we being misled?
\newblock {\em Annals of Internal Medicine}, 125:605--613, 1996.

\bibitem{Prentice1986}
RL~Prentice.
\newblock A case-cohort design for epidemiologic cohort studies and disease
  prevention trials.
\newblock {\em Biometrika}, 73:1--11, 1986.

\bibitem{BreslowHolubkov1997}
Norman~E Breslow and Richard Holubkov.
\newblock Maximum likelihood estimation of logistic regression parameters under
  two-phase, outcome-dependent sampling.
\newblock {\em Journal of the Royal Statistical Society: Series B (Statistical
  Methodology)}, 59(2):447--461, 1997.

\bibitem{RobinsGreenland1992}
JM~Robins and S~Greenland.
\newblock Identifiability and exchangeability of direct and indirect effects.
\newblock {\em Epidemiology}, 3:143--155, 1992.

\bibitem{Pearl2001}
J~Pearl.
\newblock {\em Direct and {I}ndirect {E}ffects.}
\newblock Morgan Kaufmann, San Francisco, 2001.

\bibitem{Barlowetal1999}
WE~Barlow, L~Ichikawa, D~Rosber, and S~Izumi.
\newblock Analysis of case-cohort designs.
\newblock {\em Journal of Clinical Epidemiology}, 52:1165--1172, 1999.

\bibitem{Dunningetal2015}
Andrew~J Dunning, Jennifer Kensler, Laurent Coudeville, and Fabrice Bailleux.
\newblock Some extensions in continuous models for immunological correlates of
  protection.
\newblock {\em BMC Medical Research Methodology}, 15(1):107, 2015.

\bibitem{westling2020causal}
Ted Westling, Peter Gilbert, and Marco Carone.
\newblock Causal isotonic regression.
\newblock {\em Journal of the Royal Statistical Society: Series B (Statistical
  Methodology)}, 82(3):719--747, 2020.

\bibitem{WestlingCarone2020}
T~Westling and M~Carone.
\newblock A unified study of nonparametric inference for monotone functions.
\newblock {\em Annals of Statistics}, 48(2):1001--1024, 2020.

\bibitem{PriceGilbertvanderLaan2018}
Brenda~L Price, Peter~B Gilbert, and Mark~J van~der Laan.
\newblock Estimation of the optimal surrogate based on a randomized trial.
\newblock {\em Biometrics}, 74(4):1271--1281, 2018.

\bibitem{son2020fast}
Hyunju Son and Youyi Fong.
\newblock Fast grid search and bootstrap-based inference for continuous
  two-phase polynomial regression models.
\newblock {\em Environmetrics}, in press, 2020.

\bibitem{Sachsetal2020}
J~Dudasova, L~Regina, V~Chandni, MC~Wiener, F~Gheyas, P~Fiser, J~Ivanauskaite,
  F~Liu, and JR~Sachs.
\newblock A method to estimate probability of disease and vaccine efficacy from
  clinical trial immunogenicity data.
\newblock {\em NPJ Vaccines}, 2020.

\bibitem{vanderWeeleDing2017}
TJ~VanderWeele and P~Ding.
\newblock Sensitivity analysis in observational research: introducing the
  {E}-value.
\newblock {\em Annals of Internal Medicine}, 167(4):268--74, 2017.

\bibitem{vanderWeeleMathur2020}
TJ~VanderWeele and MB~Mathur.
\newblock Commentary: developing best-practice guidelines for the reporting of
  {E}-values.
\newblock {\em International Journal of Epidemiology}, Aug 2, 2020.

\bibitem{DingvanderWeele2016}
P~Ding and TJ~VanderWeele.
\newblock Sensitivity analysis without assumptions.
\newblock {\em Epidemiology}, 27(3):368, 2016.

\bibitem{loux2017comparison}
Travis~M Loux, Christiana Drake, and Julie Smith-Gagen.
\newblock A comparison of marginal odds ratio estimators.
\newblock {\em Statistical methods in medical research}, 26(1):155--175, 2017.

\bibitem{Capedingetal2014}
Maria~Rosario Capeding, Ngoc~Huu Tran, Sri Rezeki~S Hadinegoro, Hussain Imam
  HJ~Muhammad Ismail, Tawee Chotpitayasunondh, Mary~Noreen Chua, Chan~Quang
  Luong, Kusnandi Rusmil, Dewa~Nyoman Wirawan, Revathy Nallusamy, Punnee
  Pitisuttithum, Usa Thisyakorn, In-Kyu Yoon, Diane van~der Vliet, Edith
  Langevin, Thelma Laot, Yanee Hutagalung, Carina Frago, Mark Boaz, T~Anh
  Wartel, Nadia~G Tornieporth, Melanie Saville, Alain Bouckenooghe, and the
  CYD14 Study~Group.
\newblock Clinical efficacy and safety of a novel tetravalent dengue vaccine in
  healthy children in {Asia}: a phase 3, randomised, observer-masked,
  placebo-controlled trial.
\newblock {\em The Lancet}, 384(9951):1358--1365, 2014.

\bibitem{Villaretal2015}
Luis Villar, Gustavo~Horacio Dayan, Jos{\'e}~Luis Arredondo-Garc{\'\i}a,
  Doris~Maribel Rivera, Rivaldo Cunha, Carmen Deseda, Humberto Reynales,
  Maria~Selma Costa, Javier~Osvaldo Morales-Ram{\'\i}rez, Gabriel Carrasquilla,
  Luis~Carlos Rey, Reynaldo Dietze, Kleber Luz, Enrique Rivas, Maria
  Consuelo~Miranda Montoya, Margarita~Cort\'es Supelano, Betzana Zambrano,
  Edith Langevin, Mark Boaz, Nadia Tornieporth, Melanie Saville, and {Fernando
  Noriega for the CYD15 Study Group}.
\newblock Efficacy of a tetravalent dengue vaccine in children in {L}atin
  {A}merica.
\newblock {\em New England Journal of Medicine}, 372:113--123, 2015.
\newblock DOI: 10.1056/NEJMoa1411037.

\bibitem{Vigne2017}
Claire Vigne, Martin Dupuy, Aline Richetin, Bruno Guy, Nicholas Jackson,
  Matthew Bonaparte, Branda Hu, Melanie Saville, Danaya Chansinghakul, Fernando
  Noriega, and E~Plennevaux.
\newblock Integrated immunogenicity analysis of a tetravalent dengue vaccine up
  to 4 years after vaccination.
\newblock {\em Human Vaccines \& Immunotherapeutics}, 13(9):2004--2016, 2017.

\bibitem{van2010collaborative}
Mark~J van~der Laan and Susan Gruber.
\newblock Collaborative double robust targeted maximum likelihood estimation.
\newblock {\em The International Journal of Biostatistics}, 6(1), 2010.

\bibitem{vanderWeele2008}
TJ~VanderWeele.
\newblock Simple relations between principal stratification and direct and
  indirect effects.
\newblock {\em Statistics and Probability Letters}, 78:2957--2962, 2008.

\bibitem{FDAGuidance2018}
{U.S. Department of Health and Human Services, U.S. Food and Drug
  Administration (FDA), Center for Drug Evaluation and Research (CDER), Center
  for Veterinary Medicine (CVM)}.
\newblock 2018.

\end{thebibliography}

\clearpage

\clearpage

\comment{
\begin{figure}
\begin{center}
\includegraphics[height=7in,angle=0]{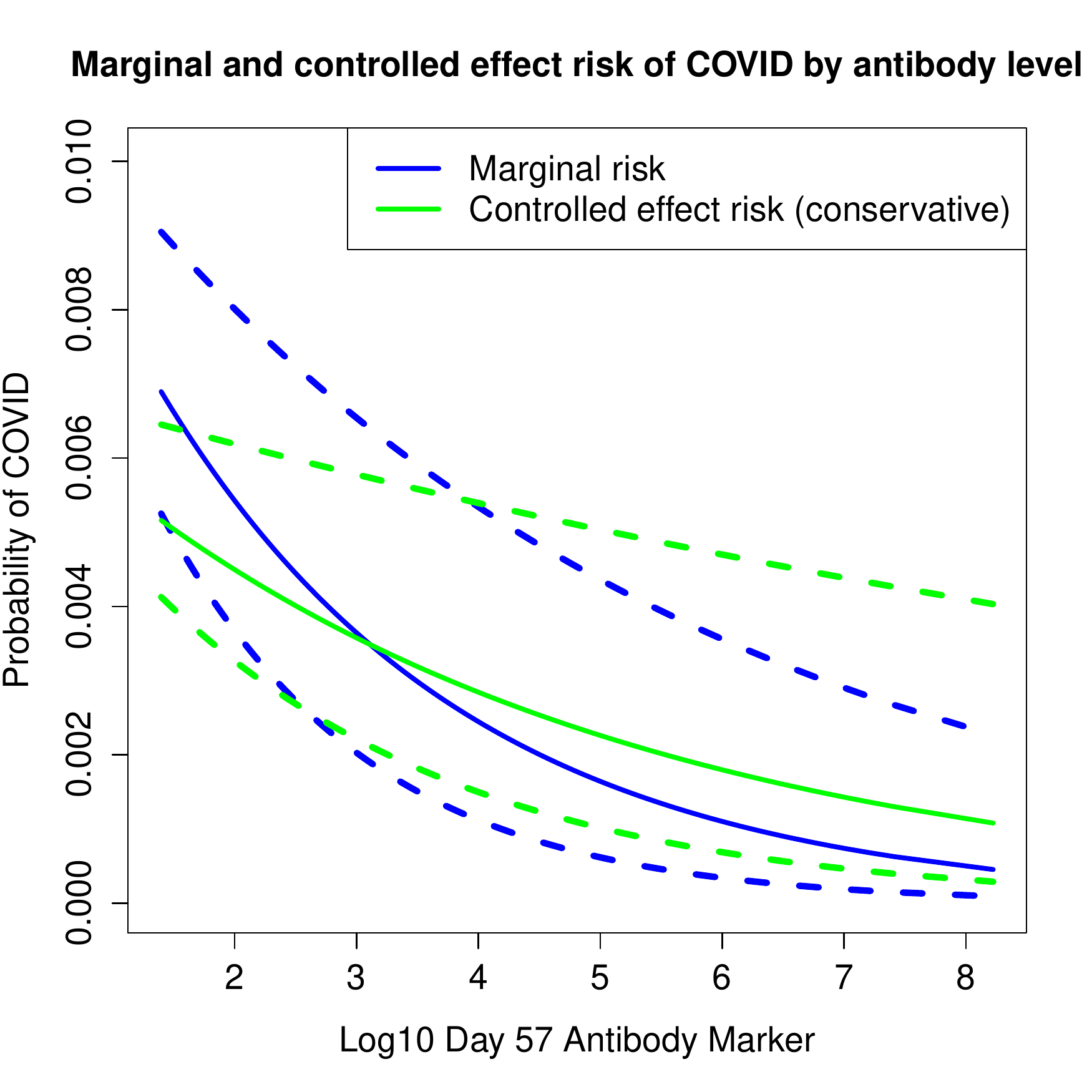}
\end{center}
\caption{Analysis of M13 average titer (quantitative) as a CoR and a controlled CoP: CYD14 and CYD15 dengue vaccine efficacy trials}\label{denguecontinuousanaysis1}
\end{figure}
}

\clearpage 
\begin{figure}
\begin{center}
\includegraphics[width=\textwidth]{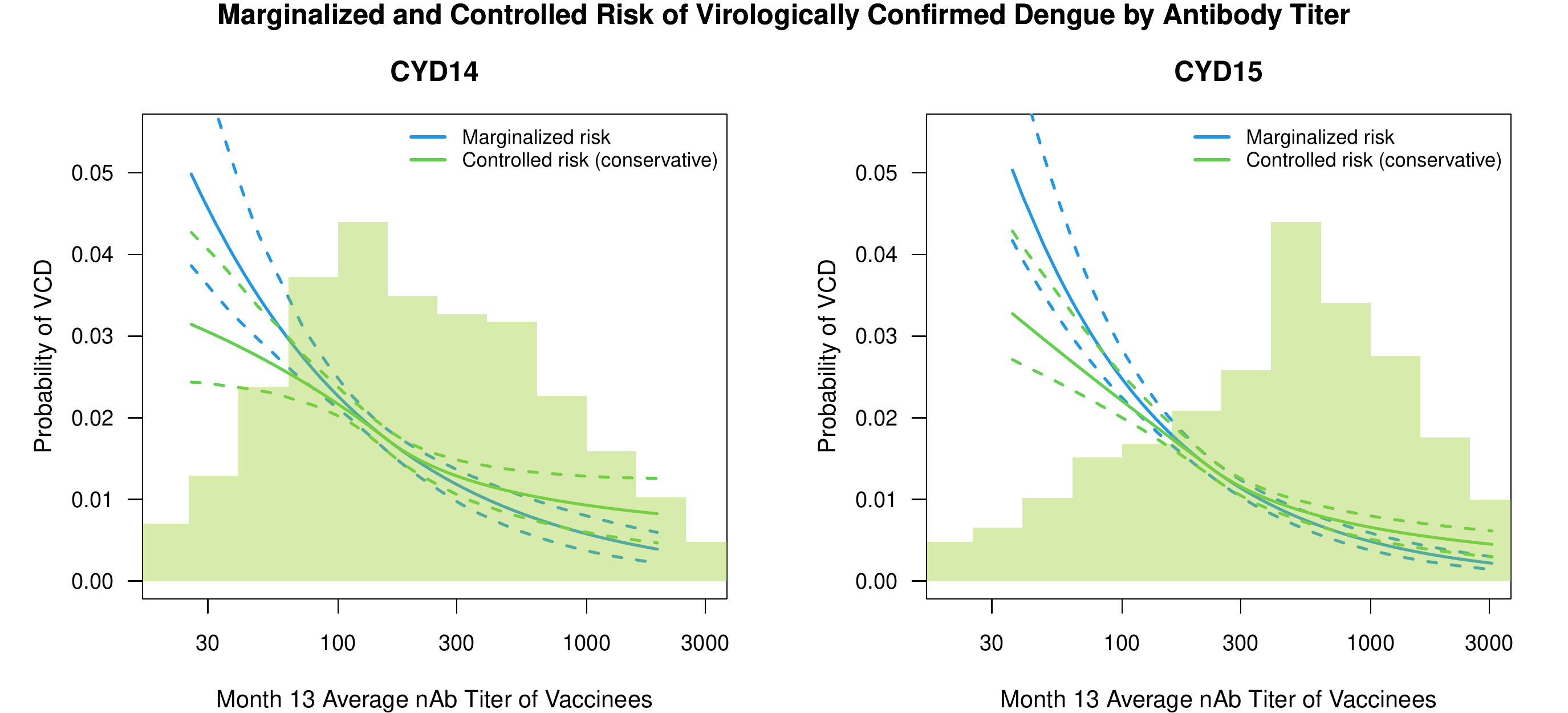}
\end{center}
\caption{Analysis of M13 average titer $S$ (quantitative) as a CoR and a controlled risk CoP: CYD14 and CYD15 dengue vaccine efficacy trials. Solid lines are point estimates and dashed lines are 95\% confidence intervals. Blue bands with a steeper shape are for marginalized risk $r_M(s)$ and green bands with a shallower shape are conservative estimates of controlled risk $r_C(s)$, with $RR_{UD}(s_{0.15},s_{0.85})=RR_{EU}(s_{0.15},s_{0.85})=4$ and hence $B(s_{0.15},s_{0.85})=16/7$, where $s_{0.15}$ and $s_{0.85}$ are 15th and 85th percentiles of $S|A=1$. 
} \label{denguecontinuousanaysis}
\end{figure}

\clearpage 

\begin{figure}
\begin{center}
\includegraphics[width=\textwidth]{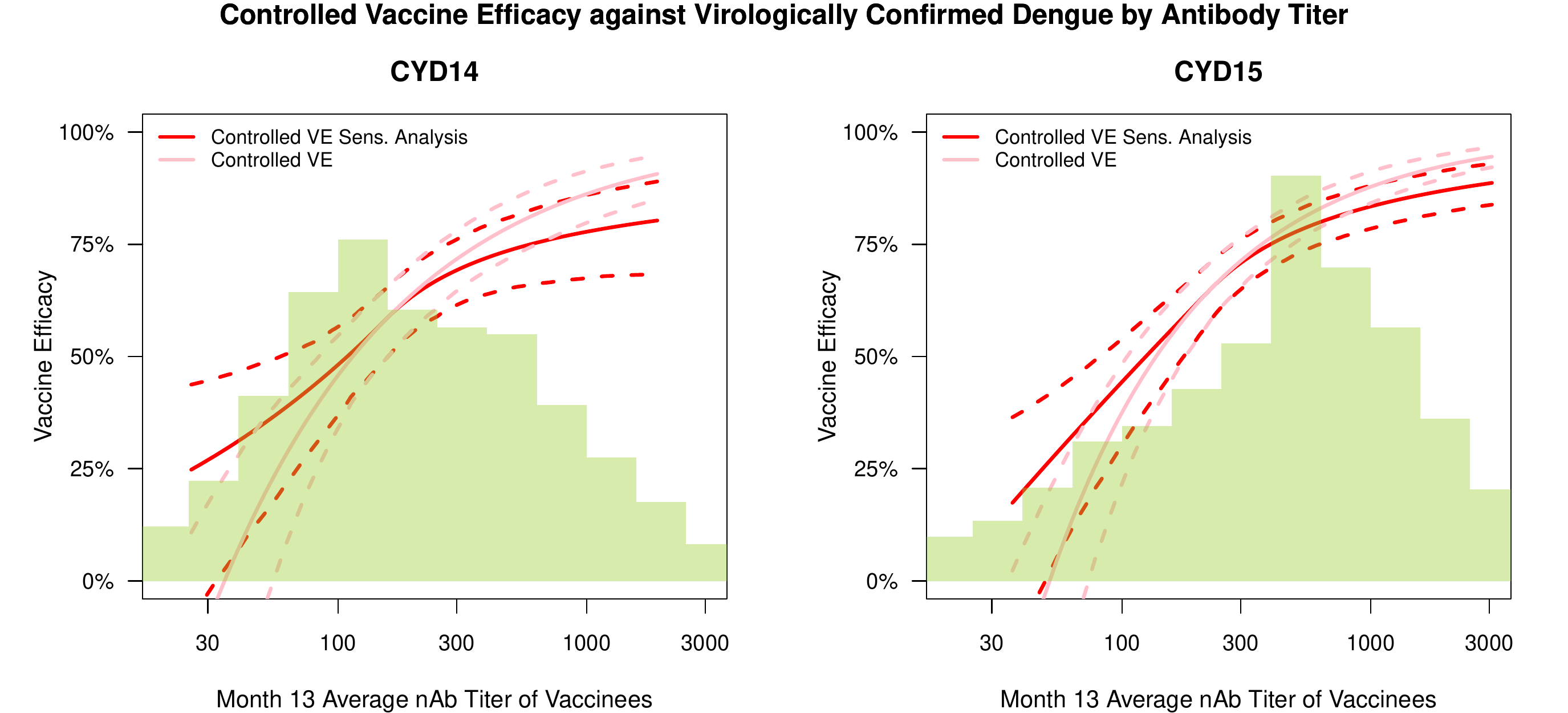}
\end{center}
\caption{Analysis of M13 average titer $S$ (quantitative) as a controlled VE CoP: CYD14 and CYD15 dengue vaccine efficacy trials. 
Solid lines are point estimates and dashed lines are 95\% confidence intervals.
The faint lines are estimates of $\mathit{CVE}(s)$ assuming no unmeasured confounding and the darker lines are conservative estimates of $CVE(s)$ accounting for potential unmeasured confounding, using the same sensitivity parameters as for Figure 2.}\label{CoPveryhighVE_Fig3}
\end{figure}

\clearpage

\end{document}